\documentclass[pra,aps,twocolumn,floats,superscriptaddress,showpacs]{revtex4}
\usepackage{psfig}
\usepackage{bm}

\begin{document}

\title{Conditional generation of arbitrary multimode 
entangled states of light \\ with linear optics}

\author{J. Fiur\'{a}\v{s}ek}
\affiliation{Ecole Polytechnique, CP 165, 
Universit\'{e} Libre de Bruxelles, 1050 Brussels, Belgium }
\affiliation{Department of Optics, Palack\'{y} University, 
17. listopadu 50, 77200 Olomouc, Czech Republic}

\author{S. Massar}
\affiliation{Ecole Polytechnique, CP 165, 
Universit\'{e} Libre de Bruxelles, 1050 Brussels, Belgium }
\affiliation{Service de Physique Th\'{e}orique, CP 225, 
Universit\'{e} Libre de Bruxelles, 1050 Brussels, Belgium}

\author{N.J. Cerf\,}
\affiliation{Ecole Polytechnique, CP 165, 
Universit\'{e} Libre de Bruxelles, 1050 Brussels, Belgium }

\begin{abstract}
We propose a universal scheme for the probabilistic generation of an 
arbitrary multimode entangled state of light with finite
expansion in Fock basis. 
The suggested setup involves passive linear optics, single photon sources, 
strong coherent laser beams, and photodetectors with single-photon
resolution. The efficiency of this setup may be greatly enhanced if, in
addition, a quantum memory is available.
\end{abstract}

\maketitle

\section{Introduction}

The generation of nonclassical states of light is one of the primary 
research areas in quantum optics.  In particular, the preparation 
of {\em entangled states} of light has attracted a considerable amount of 
attention recently,  since these states have been
identified as a key resource for quantum information processing
\cite{Nielsen00}.  Other interesting applications of the entangled states 
include, for instance,  
ultra-high precision measurements \cite{Yurke86,Hillery93,Brif96,Dowling98} 
and  quantum optical lithography \cite{Boto00,Bjork00,DAngelo01}.

The range of interactions between light fields that are 
experimentally accessible and feasible is rather limited, 
which restricts the class of quantum states of the optical field 
that can be generated in the lab. 
However, this class  can be  significantly extended  if one considers 
probabilistic generation schemes, whose success is conditioned on the
detection of a particular outcome of a measurement performed on some
ancilla system.  Schemes for probabilistic preparation of
Fock states \cite{DAriano00},
arbitrary superpositions of Fock states of single-mode
field \cite{Vogel93,Dakna99}, and superpositions of classically distinguishable
states (Schr\"{o}dinger-cat-like states) \cite{Song90,Dakna97},
have been found. Several setups for the generation of two-mode 
$N$-photon path-entangled states have been suggested 
\cite{Lee02,Fiurasek02,Kok02,Zou02,Cerf02cat}. 
Recently, the experimental conditional preparation of a single-photon
Fock state with negative Wigner function has been reported \cite{Lvovsky01}.

However, all the above mentioned schemes are still quite restrictive 
since they are capable to prepare only single-mode states or only 
a particular class of two-mode states (such as the $N$-photon states). 
In this paper,  we design a truly {\em universal}
scheme that can be used to probabilistically generate an {\em arbitrary}
multimode entangled state of light, provided that each mode does not contain
more than $N$ photons, where $N$ is an arbitrary but finite integer. 
The resources required for the present scheme comprise
the passive linear optical elements (beam splitters and phase shifters),
single photon sources, strong coherent laser beams, and photodetectors. 
It was shown recently that these resources are sufficient for performing
universal quantum computing 
\cite{Knill01,Pittman01,Ralph02,Franson02,Pittman02,Pittman03}. 
Our proposal provides another  illustration of  the surprising 
versatility and power of such an approach, relying only on linear optics 
and single photons.  In this context it is worth mentioning that, 
very recently, Clausen {\em et al.} suggested a scheme for 
approximate  conditional implementation of general single- or multi-mode 
operators acting on states of traveling optical field \cite{Clausen03}. 
The scheme described in \cite{Clausen03} is also based on linear optics  
and may be in principle
used for (approximate) quantum state preparation. We emphasize that
the approach proposed in the present paper conceptually differs from that of 
Ref. \cite{Clausen03}. We are primarily interested in exact state preparation 
and our setup is specifically tailored for that purpose.

We will first explain  all the essential features of
the state-preparation procedure on the simplest yet
nontrivial example of two-mode entangled state that 
is formed by a  superposition of two product states,
\begin{equation}
\label{trgt}
|\psi\rangle_{AB}=\frac{1}{\sqrt{\cal{N}}}(|f\rangle_A|f'\rangle_B
+ e^{i\theta}|g\rangle_A|g'\rangle_B ), 
\label{target}
\end{equation}
where $\cal{N}$ is a normalization factor.
Then we will generalize the preparation
procedure and design a scheme for the generation
of arbitrary multimode entangled states of light.
We note in passing that even the simple entangled states (\ref{target}) 
may find applications  in the 
investigations of the foundations of quantum mechanics. For instance,
it was shown recently 
that the states of the form (\ref{target}) yield a strong
violation of  Bell-CHSH inequalities  
when Alice and Bob perform balanced homodyne measurements 
followed by an appropriate binning \cite{Wenger03}.

The paper is structured as follows. In Section II,  we propose 
a method to probabilistically generate the two-mode 
entangled state  (\ref{target})
with the help of passive linear optics, single photon sources,
strong coherent laser fields and photodetectors.  
In section III  we shall extend our scheme 
to allow for more than two modes and more than two terms in the 
superposition. We will thus design a
universal device that enables one to conditionally prepare an 
arbitrary multimode  state of light. Finally, the conclusions are 
drawn in Section IV.

\section{Conditional generation of two-mode entangled states}

Let us first note that, in order to simplify the notation, 
we shall omit the normalization factors in front
of the quantum states in what follows.  

Since the state $|\psi\rangle_{AB}$ is rather complicated, we divide 
its preparation into several steps. The main simplification stems 
from the observation that the entangled state (\ref{target}) can be 
prepared by means  of entanglement swapping
\cite{Zukowski93,Pan98,Jennewein02} if we possess the two 
three-mode states
\begin{eqnarray}
|\phi\rangle_{A}&=&|f\rangle_{A_1}|V\rangle_{A_2} +|g\rangle_{A_1}|H\rangle_{A_2},
\label{targetsp}
\\
|\phi'\rangle_{B}&=&|f'\rangle_{B_1}|V\rangle_{B_2} +|g'\rangle_{B_1}|H\rangle_{B_2},
\label{targetspprime}
\end{eqnarray}
where $|V\rangle$ and $|H\rangle$ denote the state of a
single photon in spatial mode $A_2$ (or $B_2$) that is polarized vertically or
horizontally, respectively.  As shown in Fig. 1, 
we easily obtain  the state (\ref{target}) 
from the state  $|\phi\rangle_{A}|\phi'\rangle_{B}$  if we project the single 
photons in spatial modes $A_2$ and $B_2$ onto the maximally entangled Bell state
\begin{equation}
|\Phi\rangle=|V\rangle|V\rangle+e^{-i\theta}|H\rangle|H\rangle.
\label{singlet}
\end{equation}
The generation of state (\ref{target}) thus boils down to the 
preparation of the entangled state (\ref{targetsp}).
[State (\ref{targetspprime}) can be prepared similarly]. 
This latter task will be accomplished with
the help of a quantum non-demolition (QND) measurement of a single
photon, which creates entanglement.

\begin{figure}[b]
\centerline{\psfig{figure=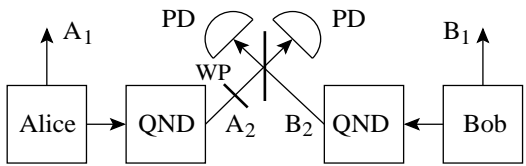,width=0.8\linewidth}}
\caption{Schematic of the setup for the
generation of the entangled state (\ref{target}) via entanglement swapping.
The two modes emerging from the QND measurement devices are combined on a 
balanced beam splitter and the swapping succeeds when each photodetector PD
detects exactly one photon. The boxes denoted Alice and Bob correspond 
to the setup depicted in Fig. 2 and $WP$ denotes a waveplate.}

\end{figure}

\subsection{QND measurement}

The proposed setup is depicted in Fig. 2. The input polarization 
modes $1V$ and $1H$ are prepared in the pure single-mode states 
$|\tilde{f}\rangle$ and $|\tilde{g}\rangle$ that contain no more than $N+1$
photons, 
\begin{equation}
|\tilde{f}\rangle=\sum_{n=0}^{N+1} \tilde{f}_{n}|n\rangle, \qquad
|\tilde{g}\rangle=\sum_{n=0}^{N+1} \tilde{g}_{n}|n\rangle,
\label{fgtildeexpansion}
\end{equation}
where $|n\rangle$ denotes the $n$-photon Fock state.
The  states (\ref{fgtildeexpansion}) are, of course, closely  related to the 
states $|f\rangle=\sum_{n=0}^N f_n|n\rangle$ and 
$|g\rangle=\sum_{n=0}^N g_n |n\rangle$  appearing in Eq. (\ref{targetsp}). 
The exact relationship between them will be specified later. 
Note that it was shown by Dakna {\em et al.} \cite{Dakna99} 
that the single-mode states (\ref{fgtildeexpansion})  
can be probabilistically generated with the help
of passive linear optics, single photon sources, strong coherent
laser pulses and photodetectors with single photon sensitivity.

\begin{figure}[!t!]
\centerline{\psfig{figure=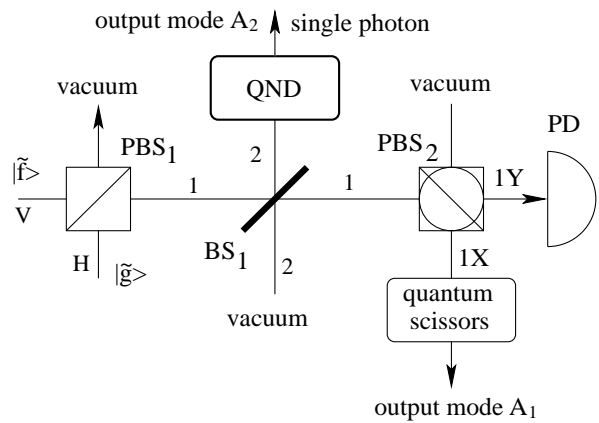,width=0.9\linewidth}}
\caption{The setup for the conditional generation of the entangled
state (\ref{targetsp}). 
The polarizing beam splitter $PBS_1$ transmits vertically
polarized photons and reflects horizontally polarized ones. The
box labeled as QND performs the quantum non-demolition measurement
of the photon number in the spatial mode while preserving the
polarization state of the photon, see Fig. 3. 
The polarizing beam splitter
$PBS_2$ partially reflects and partially transmits both vertically
and horizontally polarized photons. $BS_1$ is a usual beam splitter
and PD denotes photodetector with single-photon sensitivity.}
\end{figure}

As shown in Fig. 2, the input beam containing the states 
$|\tilde{f}\rangle$ and $|\tilde{g}\rangle$ in the $V$ and $H$ 
polarization modes impinges on a beam splitter $BS_1$ with transmittance
$t_1$ and reflectance $r_1$. The  purpose of this beam splitter is to 
separate a single photon from one of the input states and thus create 
entanglement. Of course, $BS_1$ may split more than one photon or
no photon at all. Therefore we must verify that there is exactly a single
photon present in the  output spatial mode $2$, without disturbing
its polarization state. Such QND measurements of a single photon 
have been recently thoroughly discussed in Ref. \cite{Kok02QND} 
where several schemes have been proposed and analyzed. 

One possible method relies on teleportation \cite{Bennett93,Zeilinger97}. 
With the help of the conditional C-NOT gate for photonic qubits 
\cite{Knill01} we have to first prepare the singlet state 
$|VH\rangle_{34}-|HV\rangle_{34}$ in the auxiliary modes $34$. Then we 
teleport the polarization state of a photon in mode $2$ 
onto the photon in mode $4$ by performing a Bell measurement 
on modes $2$ and $3$. This measurement should be carried out 
with two detectors that are capable  to resolve the number of photons. 
If exactly two photons are detected in coincidence at the two detectors,
then we know that there was a single photon in mode $2$ and its
polarization state has been coherently transfered onto the 
state of a single photon in mode $4$. 

Kok {\em et al.} also proposed QND measurement schemes that do not 
require a-priori entanglement \cite{Kok02QND}. 
For our purposes it suffices to employ the
simple interferometric scheme depicted in Fig. 3 which  performs a
partial QND photon number measurement. The coincident detection of a
single photon in detectors $PD_1$  and $PD_2$ indicates that there was at
least one photon in the input mode. Moreover, if there was 
{\em exactly} a single photon at the input, then its polarization state 
is unperturbed by the measurement. The probability of successful 
QND measurement of a single photon provided that there 
{\em is} a single photon in the spatial mode $2$ is equal to  $1/8$.
In the preparation scheme shown in Fig. 1 the two QND measurements are
followed by the entanglement swapping that succeeds only if each photodetector
in Fig. 1 detects exactly one photon. In this way we select the events when 
there is exactly a single photon in each mode $A_2$ and $B_2$.

The state after the QND measurement can be written in the form 
\begin{equation}
|\phi_{\rm QND}\rangle=
B_1|\tilde{f}\rangle_{1V} \, B_0|\tilde{g}\rangle_{1H} \,|V\rangle_2+
B_0|\tilde{f}\rangle_{1V} \,  B_1|\tilde{g}\rangle_{1H}\,|H\rangle_2,
\label{psiQND}
\end{equation}
where the non-unitary operators $B_0$ and $B_1$ describe removal of none
or a single photon at ${\rm BS}_1$, respectively,
\begin{eqnarray}
B_0&=&\sum_{n=0}^\infty t_1^n|n\rangle\langle n|, \nonumber \\
B_1&=&\sum_{n=0}^\infty \sqrt{n+1}\,t_1^n r_1|n\rangle\langle n+1|.
\label{B01operators}
\end{eqnarray}

\begin{figure}[t]
\centerline{\psfig{figure=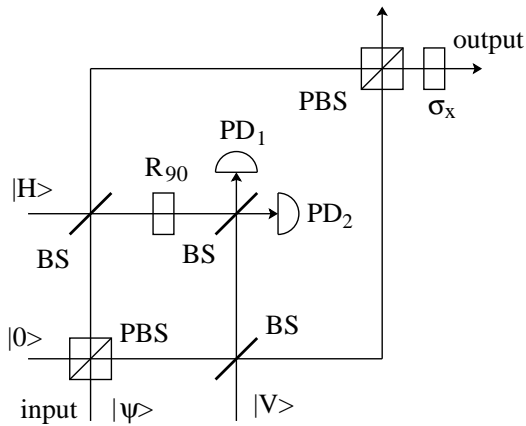,width=0.8\linewidth}}
\caption{  Interferometric scheme for a QND measurement 
of a single photon \cite{Kok02QND}. $BS$ are balanced beam splitters. $PBS$
denotes polarizing beam splitters that transmit horizontally polarized
photons and reflect vertically polarized photons. $R_{90}$ is a waveplate 
that rotates polarization state by $90$ degrees.}
\end{figure}

\subsection{Quantum erasure}

The QND measurement has thus created the entanglement. 
Still,  $B_j|\tilde{f}\rangle$  and $B_j|\tilde{g}\rangle$ are states
of two different polarization modes  while  $|f\rangle$ and
$|g\rangle$ in Eq. (\ref{targetsp}) are states of a {\em single} 
mode.  We achieve this by erasing the information about 
the polarization. This is done with the help of a polarizing 
beam splitter $PBS_2$, see Fig. 2. The beam
splitter is rotated such that it combines the $V$ and $H$ polarizations. 
 If we associate the annihilation
operators $a_{1V}$ and $a_{1H}$ with the modes $1V$ and $1H$, then the
transformation carried out by $PBS_2$ is given by
\begin{eqnarray}
a_{1X }=r_2 a_{1V} +t_2 a_{1H}, 
\nonumber \\
a_{1Y }=r_2 a_{1H} -t_2 a_{1V},
\label{PBStransform}
\end{eqnarray}
where the polarization modes $1X$ and $1Y$ 
are  {\em spatially separated} 
at the output of $PBS_2$ as indicated in Fig. 2. 
The erasing succeeds if and only if 
the detector $PD$ placed on the $1Y$ mode
does not detect any photon, which ensures that all 
photons have been transmitted to the linearly polarized output 
mode $1X$.

Since the input states do not contain more than $N+1$ photons each and
since a single photon was subtracted on $BS_1$, the state after 
erasure of the polarization information reads
\begin{equation}
|\phi_{\rm PBS}\rangle=\sum_{n=0}^{2N+1}(f_n|n\rangle|V\rangle+g_n|n\rangle|H\rangle).
\label{psierasure}
\end{equation} 
where the output complex amplitudes $f_n$ and $g_n$ are given by the
following formula,
\begin{eqnarray}
f_n&=& \sum_{k=0}^{n} \sqrt{k+1}\sqrt{n \choose k}
t_1^n r_1 r_2^k t_2^{n-k}\tilde{f}_{k+1}\tilde{g}_{n-k} 
, \nonumber \\
g_n&=&\sum_{k=0}^{n} \sqrt{k+1}\sqrt{n \choose k}
t_1^n r_1  r_2^{n-k} t_2^k \, \tilde{g}_{k+1}\tilde{f}_{n-k}.
\nonumber \\
\label{fgoutexact}
\end{eqnarray}  

It follows that we can fix the values of output coefficients $f_n$ 
by properly choosing the complex amplitudes of the input states. We have
the recurrence equations
\begin{eqnarray}
\tilde{f}_{n+1}&=&\frac{f_n-F_{n}}{\sqrt{n+1} t_1^n r_1 r_2^n \tilde{g}_0},
\nonumber \\
F_n&=&\sum_{k=0}^{n-1}\sqrt{k+1}\sqrt{{n \choose k}}
t_1^n r_1 r_2^k t_2^{n-k}\tilde{f}_{k+1}\tilde{g}_{n-k},
\nonumber \\
\label{inoutsolution}
\end{eqnarray}
and a similar formula holds for $\tilde{g}_{n+1}$. The amplitudes
$\tilde{f}_{n+1}$ and $\tilde{g}_{n+1}$ are
calculated as follows. First, some nonzero values for
$\tilde{f}_0$ and $\tilde{g}_0$ are chosen. Then one uses repeatedly 
Eq. (\ref{inoutsolution})  to calculate $\tilde{f}_{n+1}$ 
and $\tilde{g}_{n+1}$ from 
$\tilde{f}_k$ and $\tilde{g}_k$, $0\leq k\leq n$,
and the iterations stop at $n=N$.

We have thus almost obtained the  state (\ref{targetsp}). However, while
the effective states $|f\rangle$ and $|g\rangle$ in (\ref{psierasure}) 
have the correct structure in the subspace spanned by the first $N+1$ 
Fock states $|0\rangle,|1\rangle,\ldots, |N\rangle$, 
they also contain an unwanted tail in the subspace spanned 
by the Fock states $|N+1\rangle, \ldots, |2N+1\rangle$.

\begin{figure}
\centerline{\psfig{figure=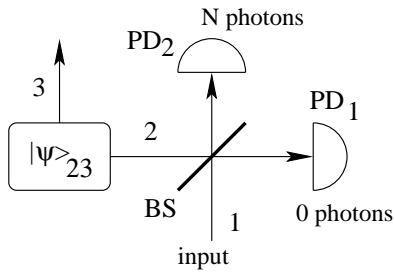,width=0.6\linewidth}}
\caption{Quantum scissors.}
\end{figure}

\subsection{Quantum state truncation}

The last step in our scheme is to get rid of that tail. We do so with the
help of the quantum scissors
\cite{Pegg98,Koniorczyk00,VillasBoas01,Ozdemir01,Babichev02} 
that will project the state in
the mode $1$ onto the subspace spanned by the first $N+1$ Fock states. This
action is formally described by the projector $\Pi_{QS}=\sum_{n=0}^N
|n\rangle\langle n|$. After this truncation, we finally obtain the target
entangled state (\ref{targetsp}) in the output modes 1 and 2 of our device.

The quantum scissors have been discussed in detail in several recent
papers \cite{Pegg98,Koniorczyk00,VillasBoas01,Ozdemir01,Babichev02}.
The schemes proposed there can be  implemented with the use of
linear optics, single photon sources, and photodetectors. 
The setup that we shall consider in the present paper is depicted 
in Fig. 4. The modes $2$ and $3$ are prepared in a pure $N$-photon two-mode entangled state 
\begin{equation}
|\psi\rangle_{23}=\sum_{k=0}^N c_k|k\rangle_2|N-k\rangle_3.
\label{psiscissors}
\end{equation}
It was shown recently that these states may be
conditionally generated with the help of linear optics, single photon 
sources and photodetectors \cite{Lee02,Fiurasek02,Kok02,Zou02}. 
We then combine the modes $1$ and $2$ on a balanced beam splitter and
measure the number of photons in output modes $1$ and $2$ by means of 
two photodetectors $PD_1$ and $PD_2$ with single-photon resolution. 
The truncation is a conditional
operation that succeeds when certain particular measurement outcome has
been detected, say, no photons at $PD_1$ and $N$ photons at $PD_2$, which
we denote as $(0,N)$. Note,
that one could also consider other detection events, but the total 
number of photons detected by the photodetectors should be equal to $N$.
The quantum scissors is essentially a quantum state teleportation in the
subspace of the first $N+1$ Fock states, where the state 
(\ref{psiscissors}) serves as the
quantum channel, and the Bell-type measurement is carried out with the use
of the beam splitter in Fig. 4 that couples the modes $1$ and $2$.

For our particular choice of the detection event $(0,N)$, and assuming a
balanced beam splitter, we find that 
\begin{equation}
c_k=\sqrt{P_{\rm QS}} \frac{\sqrt{k!(N-k)!}}{2^{-N/2}\, \sqrt{N!}},
\label{ck}
\end{equation}
should hold, where
\begin{equation}
P_{\rm QS}=\left(\sum_{k=0}^N \frac{ k! (N-k)!}{2^{-N \,}N!}\right)^{-1}
\label{PNQS}
\end{equation}
is the probability of the successful quantum-state truncation that projects
onto the subspace spanned by $|0\rangle,|1\rangle,\ldots,|N\rangle$.

\section{General scheme}

So far we have considered the generation of a two-mode state that 
consists of a superposition of two product states. In this section we shall
generalize our scheme in several ways.

\subsection{$M$-mode entangled state}
First, we will
discuss the preparation of a $M$-mode entangled state which has a form 
limited to the superposition of two product states, namely
\begin{equation}
|\psi\rangle=|f_1\rangle |f_2\rangle  \ldots |f_M\rangle
+e^{i\theta}|g_1\rangle|g_2\rangle
\ldots|g_M\rangle.
\label{psiNmode}
\end{equation}
To prepare this state, we first generate $M$ states 
\begin{equation}
|\phi_j\rangle=|f_j\rangle|V\rangle+|g_j\rangle|H\rangle
\label{psispNmode}
\end{equation}
as before, and then project the $M$ single photons onto 
the maximally entangled state
\begin{equation}
|GHZ\rangle=|VV\ldots V\rangle+e^{-i\theta}|HH\ldots H\rangle.
\label{GHZstate}
\end{equation}
This projection  can be performed with the  
Greenberger-Horne-Zeilinger (GHZ) state analyzer described 
by Pan and Zeilinger \cite{Pan98GHZ}. 
This analyzer consists of polarizing beam splitters and half-wave plates
and allows one to distinguish two GHZ states of M qubits. 

\begin{figure}[t]
\centerline{\psfig{figure=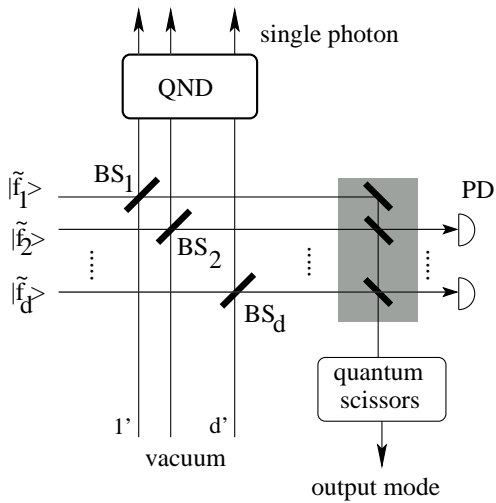,width=0.75\linewidth}}
\caption{Scheme for conditional generation of the entangled state
(\ref{psiquditsp}). The erasing of the spatial information is achieved
conditionally on no photon being detected at the $d-1$ photodetectors $PD$.}
\end{figure}

\subsection{Arbitrary two-mode entangled states}

Our second generalization extends the preparation procedure to two-mode 
states that are superpositions of $d$ product states,
\begin{equation}
|\psi\rangle_{AB}=\sum_{j=1}^d e^{i\theta_j} |f_{j}\rangle_A|g_{j}\rangle_B.
\label{psiqudit}
\end{equation}
Obviously, a natural strategy is to prepare this state via
$d$-dimensional entanglement swapping from two states of the form 
\begin{equation}
|\phi\rangle_{AC}=\sum_{j=1}^d |f_j\rangle_{A}|1\rangle_{jC},
\qquad 
|\phi\rangle_{BD}=\sum_{j=1}^d |g_j\rangle_{B}|1\rangle_{jD},
\label{psiquditsp}
\end{equation}
where $|1\rangle_{j}$ denotes a state of a single photon in the $j$th auxiliary
spatial mode.  To prepare the states (\ref{psiquditsp}), 
we proceed essentially as in
Sec. II. The extended scheme is depicted in Fig. 5.  Initially, $d$ spatial
modes are prepared in states $|\tilde{f}_j\rangle$. The array of $d$ beam
splitters $BS_j$ extracts, with a certain probability, a single photon.  
We must verify the presence of exactly one single photon in total in
the modes $1',\ldots, d'$ while preserving the coherence among these modes
 by performing a QND measurement. This can be 
accomplished by $d$-dimensional teleportation similarly 
as discussed in Section II.  However, this time
we need to teleport a qudit encoded as a state of a single photon in 
$d$ spatial modes. Fortunately, it was shown recently that the
probabilistic gates for quantum computing with linear optics 
\cite{Knill01,Pittman01,Ralph02,Franson02,Pittman02}, that were
introduced for qubits, can be easily extended to qudits
\cite{Dusek01}. This means that all the necessary manipulations, such as
the generation of maximally entangled state of two qudits and the Bell
measurement, can be probabilistically carried out  with the resources 
that we assume here.
 
After the QND measurement, we must erase the spatial information contained
in the resulting state 
\begin{equation}
|\psi_{QND}\rangle=\sum_{j=1}^d \left(\prod_{k=1}^d B_{\delta_{j,k}}|\tilde{f}_k\rangle
\right)\,|1\rangle_j, 
\end{equation}
where $B_j$ are given by Eq. (\ref{B01operators}). 
This is done by the second array of $d-1$ beam splitters indicated by
the shaded zone in Fig. 5.
The preparation is finished by the application of the quantum scissors,
that truncate the Fock state expansion of the output state at $N$ photons.

Suppose that the output mode after erasing is a balanced linear superposition
of all $d$ modes
\begin{equation}
a_{\rm out}=\frac{1}{\sqrt{d}}\sum_{j=1}^d a_j.
\end{equation}
The complex amplitudes of the states $|f_j\rangle=\sum_n f_{j,n}|n\rangle$
can then be expressed in terms of the complex amplitudes of the input states 
$\tilde{f}_{j,n}$ as follows
\begin{equation}
f_{j,n}=\sqrt{n!}\,
 \frac{t_1^n r_1}{d^{n/2}} \,
{ \sum_{\bm{n}}}^\prime \sqrt{n_j+1}\frac{\tilde{f}_{j,n_j+1}}{\sqrt{n_j!}}
\prod_{k\neq j} \frac{\tilde{f}_{k,n_k}}{\sqrt{n_k!}},
\label{finoutglobal}
\end{equation}
where the prime indicates summation over all $\bm{n}=(n_1,\ldots,n_d)$ 
satisfying the constraint $\sum_{k=1}^d n_k=n$. 
Note that the formula (\ref{finoutglobal}) may be easily inverted and
we can determine the complex amplitudes $\tilde{f}_{j,k}$ of the input states
for any given prescribed output states $|f_j\rangle$.

\subsection{Universal scheme}
 
Finally,  we point out that those two  generalizations can 
be combined and we can thus generate an arbitrary multimode entangled 
state of light where each mode contains no more than $N$ photons. 
Any $M$-mode state of this kind can be written as a superposition of no more
than $d=(N+1)^M$ product states,
\begin{equation}
|\psi\rangle=\sum_{j=1}^d  e^{i\theta_j}
|f_{j1}\rangle_1|f_{j2}\rangle_2\ldots|f_{jM}\rangle_M.
\label{superposition}
\end{equation}
Note that the norm of each  product state in the superposition 
(\ref{superposition})  may be arbitrary.
We can conditionally generate the 
state (\ref{superposition}) if we first prepare $M$ entangled states
($k=1,\ldots,M$)
\begin{equation}
|\phi_{k}\rangle= \sum_{j=1}^d  |f_{jk}\rangle |1\rangle_{jk},
\label{superpositionsp}
\end{equation}
where $|1\rangle_{jk}$ denotes a single photon state 
of the $(j,k)$-th auxiliary spatial mode. As discussed above, 
the states (\ref{superpositionsp}) can be prepared  
with the help of the scheme depicted in Fig. 5. 
Then, 
we carry out a joint measurement on all auxiliary photons. 
The state $|\psi\rangle$ is obtained if we project the auxiliary 
modes onto an $M$-photon GHZ entangled state:
\begin{equation}
|GHZ_{Md}\rangle=\sum_{j=1}^d  e^{-i\theta_j} |1\rangle_{j1}|1\rangle_{j2}
\ldots|1\rangle_{jM}.
\label{GHZM}
\end{equation}
If we denote 
$|\Phi_{Md}\rangle=|\phi_1\rangle|\phi_2\rangle\ldots|\phi_M\rangle$
then  we can  write 
\begin{equation}
|\psi\rangle=\langle GHZ_{Md}|\Phi_{Md}\rangle.
\end{equation}

With the help of the techniques developed in the framework of the quantum
computation  with linear optics, the entangled $M$-photon GHZ
state  (\ref{GHZM}) can be mapped onto 
a product state of $M$ photons which can then be detected by observing an
appropriate $M$-photon coincidence. 
The factorization  consists of a sequence of  the C-SHIFT gates 
for two qudits. The C-SHIFT gate is defined as follows,
\begin{equation}
|1\rangle_{jk}|1\rangle_{j^\prime k^\prime}\rightarrow 
|1\rangle_{jk} |1\rangle_{j^\prime-j,k^\prime},
\end{equation}
where $j^\prime-j$ should be calculated modulo $d$.
As shown in Ref. \cite{Dusek01} this transformation 
can be (probabilistically) performed with the use of single photon sources, 
passive linear optics and photodetectors with single photon sensitivity.
If we apply the C-SHIFT gate to the qudits $1$ and $k$, where $k=2,\ldots,M$,
then the GHZ state (\ref{GHZM}) is mapped onto product state of $M$ photons
\begin{equation}
|GHZ_{Md}\rangle \rightarrow \left(\sum_{j=1}^d e^{-i\theta_j}
|1\rangle_{j1} \right)
|1\rangle_{d2}\ldots |1\rangle_{dM}.
\end{equation}

\subsection{State preparation with quantum memory}

The above described procedure for the preparation of the generic $M$-mode 
entangled state clearly shows that our approach is indeed general since it
allows us, in principle, to generate an arbitrary state. But the success 
probability of the scheme will be exponentially small in general. 
The preparation procedure will therefore have to be repeated exponentially 
many times to produce the desired state. In the rest of this section we shall 
argue that the number of operations required to produce the desired state can be
significantly decreased if a quantum memory is available. The situation is
similar to that analyzed in Ref. \cite{Cerf02cat} where the operations required
to produce ``cat states'' was drastically decreased by using a
recursive procedure that uses a quantum memory.

The advantage of the quantum memory is that it allows one to wait until
successive steps in the state generation protocol have succeeded before
proceeding with the next steps. For instance in the scheme described in Fig. 5, 
one can first produce the states $|\tilde{f}_j\rangle$ and store them in a
memory before proceeding with the production of the entangled state 
(\ref{psiquditsp}).
Suppose that the state $|\tilde{f}\rangle_j$ can be prepared with the
probability $P_j$. Without quantum memory the probability of simultaneous
preparation of the 
$d$ states $|\tilde{f}\rangle_j$ is $P=\prod_{j=1}^d P_j$ and the
number of required operations thus scales as $NO\propto 1/P$. 
If, after a successful
preparation, we store each state $|\tilde{f}\rangle$ in a memory, we reduce the
number of required operations to $NO\propto \sum_{j=1}^d 1/P_j$.

A more elaborate use of a quantum memory is to build the state using 
a {\em recursive
procedure} based on the Schmidt decomposition.
The method is best explained on an
explicit example. Suppose we would like to prepare a four-mode state
$|\psi\rangle_{ABCD}$. The Schmidt decomposition of this four-mode state 
with respect to  a bi-partite splitting into the $AB$ and $CD$  modes can be 
written as follows,
\begin{equation}
|\psi\rangle_{ABCD}=\sum_{j=1}^d 
|\varphi_j\rangle_{AB}\,|\pi_j\rangle_{CD} \,,
\label{Schmidt}
\end{equation}
where $|\varphi_j\rangle_{AB}$ and 
$|\pi_j\rangle_{CD}$ form orthogonal bases 
in some subspaces of the  Hilbert space of modes $AB$ and $CD$, respectively.
The number of the terms in the Schmidt
decomposition (\ref{Schmidt}) is bounded by $d\leq (N+1)^2$ while, in general,
$(N+1)^4$ terms were necessary in the general procedure of Section IIIC.
The decomposition (\ref{Schmidt}) suggests that we could obtain the state
$|\psi\rangle_{ABCD}$ via entanglement swapping if we first prepare two
entangled states
\begin{eqnarray}
|\Phi\rangle_{ABX}=\sum_{j=1}^d |{\varphi_j}\rangle_{AB}|1\rangle_{jX},
\nonumber \\
|\Phi'\rangle_{CDY}=\sum_{j=1}^d | \pi_j\rangle_{CD}|1\rangle_{jY}.
\label{ABX}
\end{eqnarray}
Now the entangled states $|\Phi\rangle_{ABX}$ and $|\Phi^\prime\rangle_{CDY}$
can be prepared using the general
procedure of Sec. IIIC. However, for states of the specific form (\ref{ABX}),
a simpler procedure can be devised. 
One  first generates $d$ states $|\tilde{\varphi}_j\rangle_{AB}$ where the
tilde indicates that $|\tilde{\varphi}_j\rangle$ is related to 
$|\varphi_j\rangle$ via relations similar to those described by Eqs. 
(\ref{inoutsolution}). The state $|\Phi\rangle_{ABX}$ is then prepared with
the help of the device shown in Fig. 5 where 
the mode $B$ of $|\tilde{\varphi}_j\rangle_{AB}$
is sent to the $j$th input port of the interferometer.
The only modification is that we must also erase the spatial information 
carried by the $A$ modes of the states $|\tilde{\varphi}_j\rangle_{AB}$
and perform the quantum state truncation after the erasing.
Briefly, all $d$  modes  $A_j$ are combined on an array of $d-1$ 
beam splitters and the detectors monitor the first $d-1$ output ports. 
If these detectors do not  register any photon, then the erasing procedure 
succeeded and we apply the quantum scissors to the state in the $d$th output
port.  

Clearly, this recursive procedure can be extended to any number of modes.
The resulting scheme resembles a tree structure involving repeated 
applications of the entangling scheme shown in  Fig. 5 and the entanglement
swapping that produces an $2m$-mode state from two $m$-mode states  
entangled with auxiliary single photons. 
This scheme relying on quantum memory may possibly lead to a substantial 
improvement of the generation  probability,  similarly to the case 
of the two-mode $N$-photon  cat states \cite{Cerf02cat}.

\section{Conclusions}

In summary, the scheme proposed in the present paper 
is universal in that arbitrary multimode entangled
states of light can be probabilistically generated. The resources
required are passive linear optics, single photon sources, 
strong coherent states, and detectors with single-photon resolution. 
This result clearly illustrates the versatility and power of the approach
relying only on linear optics and single photons. 

It is fair to say, though, that the suggested setup is rather
complicated and involves several nontrivial operations such as the quantum
non-demolition measurement of a single photon, quantum scissors, and
transformations of photonic qudits encoded as a state of a single photon in
$d$ spatial modes. We have shown that all these operations can be
(probabilistically) implemented with the resources that we consider here.
The resulting procedure may become very complex for general states.
However, it should be emphasized that
 building a state like (\ref{superposition}) must necessarily
be complicated
because of the large number of parameters [$(N+1)^M-1$ complex numbers] that
characterize this state and that must be fixed by the preparation scheme.
We have also argued that the use of a quantum memory can decrease the number of
operations required to produce the desired state.

We hope that this paper will stimulate the efforts towards experimental
demonstrations of the basic building blocks of our scheme, 
such as the conditional generation of single-mode finite superpositions 
of Fock states and the preparation of two-mode entangled $N$-photon 
states required for the quantum state truncation.

\acknowledgments
 
We are grateful to M. Du\v{s}ek, R. Filip and P. Grangier 
for stimulating discussions.
We acknowledge financial support from the Communaut\'e Fran\c{c}aise de
Belgique under grant ARC 00/05-251, from the IUAP programme of the Belgian
government under grant V-18, from the EU under project RESQ
(IST-2001-37559) and CHIC (IST-2001-32150). 
J.F. was also partially supported by 
the grant LN00A015 of the Czech Ministry of Education.

\end{document}